\documentclass[preprint,prb,a4paper,showpacs,superscriptaddress,showkeys]{revtex4}

\usepackage{latexsym}
\usepackage{amsmath}
\usepackage{dcolumn}

\usepackage{color}

\usepackage[dvips]{graphicx}
\graphicspath{{../figs/}}
\usepackage{subfigure}

\usepackage[utf8x,latin1]{inputenc}

\newcommand {\qref}[1]{Ref.~\onlinecite{#1}}

\newcommand {\qsect}[1]{Sect.~\ref{#1}}

\newcommand {\queq}[1]{(\ref{#1})}
\newcommand {\qeq}[1]{Eq.~\queq{#1}}

\newcommand {\Om}{\mathbf{\Omega}}

\newcommand {\Er}{E_r}
\newcommand {\Erfit}{E_r^{\rm fit}}
\newcommand {\eVA}{eV/\AA$^3$}
\newcommand {\Dt}{\Delta t}
\newcommand {\ani}{A}
\newcommand {\eps}{\epsilon}

\newcommand {\gamu}{\gamma_u}
\newcommand {\gams}{\gamma_s}
\newcommand {\Gr}{{G_r}'}

\newcommand {\slip}{$\langle 1\bar{1}0\rangle$ \{111\}}
\newcommand {\two}{$\langle 11\bar{2}\rangle$ \{111\}}
\newcommand {\szero}{s_{100}}
\newcommand {\sone}{s_{111}}

\newcommand {\Ecoh}{E_{\rm coh}}
\newcommand {\sigcr}{\sigma_{\rm cr}}
\newcommand {\tauth}{\tau_{\rm th}}
\newcommand {\dy}{d_{\rm yield}}
\newcommand {\rcut}{r_{\rm cut}}
\newcommand {\VLJ}{V_{\rm LJ}}
\newcommand {\VMorse}{V_{\rm Morse}}
\newcommand {\VEAM}{V_{\rm EAM}}

\begin{document}

\date{\today}

\title{Pair vs many-body potentials: influence on elastic and plastic
behavior in nanoindentation
}

\author{Gerolf Ziegenhain}
\affiliation{%
Fachbereich Physik und Forschungszentrum OPTIMAS,
Universit{\"a}t Kaiserslautern, \\
Erwin-Schr{\"o}dinger-Stra{\ss}e, D-67663 Kaiserslautern, Germany}

\author{Alexander Hartmaier}
\affiliation{%
Interdisciplinary Center for Advanced Materials Simulation (ICAMS),
Ruhr-University Bochum,\\ Stiepeler Str.\ 129 (UHW), D-44780 Bochum,
Germany}

 \author{Herbert M.~Urbassek}
 \email{urbassek@rhrk.uni-kl.de}
 \homepage{ http://www.physik.uni-kl.de/urbassek/}
\affiliation{%
Fachbereich Physik und Forschungszentrum OPTIMAS,
Universit{\"a}t Kaiserslautern, \\
Erwin-Schr{\"o}dinger-Stra{\ss}e, D-67663 Kaiserslautern, Germany}

\begin{abstract}

Molecular-dynamics simulation can give atomistic information on the
processes occurring in nanoindentation experiments. In particular, the
nucleation of dislocation loops, their growth, interaction and motion
can be studied. We investigate how realistic the interatomic potentials
underlying the simulations have to be in order to describe these complex
processes. Specifically we investigate nanoindentation into a Cu single
crystal. We compare simulations based on a realistic many-body
interaction potential of the embedded-atom-method type with two simple
pair potentials, a Lennard-Jones and a Morse potential. We find that
qualitatively many aspects of nanoindentation are fairly well reproduced
by the simple pair potentials: elastic regime, critical stress and
indentation depth for yielding, dependence on the crystal orientation,
and even the level of the hardness. The quantitative deficits of the
pair potential predictions can be traced back (i) to the fact that the
pair potentials are unable in principle to model the elastic anisotropy
of cubic crystals; (ii) as the major drawback of pair potentials we
identify the gross underestimation of the stable stacking fault energy.
As a consequence these potentials predict the formation of too large
dislocation loops, the too rapid expansion of partials, too little cross
slip and in consequence a severe overestimation of work hardening.

\end{abstract}


\maketitle

%\newpage

\section{Introduction}

Nanoindentation into crystalline materials is a complex
process.\cite{Fis04,GCD*07} When the indenter is moved into the surface,
it deforms the substrate first elastically, until a sufficiently high
pressure
has been established and plasticity sets in. This so-called `critical
indentation depth' is characterized by the critical stress necessary for
the nucleation of dislocations. Upon further indentation, the emerging
plasticity will initially lead to a drop in the contact pressure felt by
the indenter -- the `load drop' -- but then the contact pressure will
saturate; its value is then called the hardness of the material. As soon
as dislocations have been generated, they will propagate, multiply,
cross-slip, interact with each other, etc. The multitude of the
processes which these dislocations undergo will eventually have a back
reaction on the proceeding indenter: the indenter will not penetrate
into virgin material but into work hardened material.

Molecular-dynamics simulation has been employed to obtain a detailed
in-depth understanding of the processes occurring during
nanoindentation, and in particular in the plastic
regime.\cite{LLBC90,KPH98,SCK*03,VLZ*03,Li07,Szl06,GCD*07} The advantage
of this method is the detailed atomistic information it can provide on
virtually all the processes occurring in the material, and in fact,
since the 1990s, an increasing number of simulations have been performed
and contributed to our understanding of nanoindentation and plasticity
in general. The use of these simulations is impeded by the large
simulation volumes which are necessary to host the defects formed, and
the long time scales over which simulations need to be followed. Besides
these problems, in principle, molecular-dynamics simulation allow a
realistic simulation of the events as long as the interatomic
interaction potential has been realistically chosen.

One of the objectives of understanding the physics of nanoindentation is
to trace back the origin of the physical phenomena observed to the
peculiarities of the interatomic interaction for the specific material
under investigation. The question then rises which feature of the
interatomic interaction potential is responsible for which phenomenon
observed in the simulation. Such questions can usually be answered with
the help of sensitivity studies, in which one or several features of the
interatomic potential are systematically varied and their effect on the
simulation results is studied. Unfortunately, realistic interaction
potentials, such as they are used nowadays, do not allow for such
systematic changes, since they are available either in the form of
parameterized analytical formulae, in which the change of one parameter
affects several physical material properties simultaneously, or even
only in the form of numerical tables. Therefore, simpler generic
potentials may be employed which allow the typical behavior of solids to
be studied without reproducing too well the specific properties of one
particular material.

In particular, pair potentials have been used to model nanoindentation
of materials or the related phenomena of plasticity, work hardening and
material
failure.\cite{HDH*90,HVW*91,KHH*93,AWG*02,AWG*02a,MY03,BHG*04,BHD*05,SF05}
It has been known
for long that pair potentials are capable to model rare gas solids, but
they have several deficits in modelling metallic materials. Thus they
allow to prescribe only two -- rather than three -- elastic constants
for solids, they model an outwards -- rather than an inwards -- surface
relaxation, etc.\cite{Car90,DFB93} However, how well are these pair
potentials able to model nanoindentation? In other words, which aspects
in the elastic and plastic deformation of the material, in the
nucleation and glide of dislocations are described qualitatively
correctly, and which are not? How large will quantitative deviations
between the predictions of a pair potential and that of realistic
potentials be?

In the present paper, we wish to answer these questions for the specific
case of nanoindentation into a Cu single crystal. For this material, a
many-body potential is available, which is well characterized also with
respect to the prediction of the mechanical properties, enjoys a rather
wide acceptance in the community, and which has been repeatedly used for
nanoindentation and plasticity simulations in the past. We shall use
simulations with this potential as a reference case and contrast the
results obtained with those predicted by using two simple pair
potentials.

\section{Method}

\subsection{Potentials}

We chose the potential developed by Mishin et al.\cite{MMP*01} as the
state-of-the-art reference potential for Cu; this potential has been
often employed for molecular-dynamics simulations and has been well
characterized.\cite{ZLV*04,TSM07,TM08,LGSC08} This potential belongs to
the class of embedded-atom-method (EAM) potentials,\cite{DB84,DFB93}
which incorporate many-body bonding effects in an appropriate form to
describe metallic bonding. In the embedded-atom method, the total energy
of a system is represented as

\begin{equation}  \label{EAM}
 E_{\rm tot} = \sum_{\substack
 {i, j \\ i \ne j }}
 \VEAM(r_{ij}) + \sum_i F(\rho_i) ,
\end{equation}

where $\VEAM(r_{ij})$ is a pair potential evaluated as a function of the
distance $r_{ij}$ between atoms $i$ and $j$, and $F$ is the embedding
energy, which depends on the so-called 'electron density' $\rho_i$. The
latter is given by

\begin{equation}  \label{rho}
 \rho_i= \sum_{j \ne i} g(r_{ij}),
\end{equation}

where $g(r_{ij})$ is the contribution of atom $j$ to the total electron
density at the site of atom $i$. The detailed form of the functions
$\VEAM$, $F$, and $g$ is presented in \qref{MMP*01}. We collect in
Table~I a number of basic characteristics of crystalline Cu: the
cohesive energy $\Ecoh$, the lattice constant $a$, the bulk modulus $B$,
and the three elastic moduli $c_{11}$, $c_{12}$ and $c_{44}$. These
properties are well reproduced by the Mishin potential.

We employ two well known pair potentials, the Morse and the
Lennard-Jones (LJ) potential. The Morse potential

\begin{equation}  \label{Morse}
\VMorse(r)=D
\left[ e^{-2\alpha(r-r_0)} - 2e^{-\alpha
(r-r_0)} \right]
\end{equation}

is characterized by three parameters: the bond strength $D$, the
equilibrium bond distance $r_0$, and the potential fall-off $\alpha$. We
fit these parameters to three materials properties; these are
traditionally chosen as the lattice constant $a$, the cohesive energy
$\Ecoh$, and the bulk modulus $B$. The latter is given in terms of the
elastic moduli by

\begin{equation}  \label{B}
B = \frac{c_{11}+ 2  c_{12}}{3} .
\end{equation}

For Cu, the parameters read $D= 0.3282$ eV, $r_0 = 2.8985$ \AA, and
$\alpha= 1.3123$ \AA$^{-1}$. As Table~I, demonstrates, the two elastic
constants $c_{11}$ and $c_{12}$ are reproduced rather well, within a 2
\% error margin. Of course, since pair potentials obey the Cauchy
relationship,\cite{Car90} $c_{44}=c_{12}$, it is not possible to fit
$c_{44}$ separately, and indeed the shear modulus is misrepresented by
almost 60 \%.

For the LJ potential,

\begin{equation}  \label{LJ}
\VLJ(r) = 4\epsilon \left[ \left( \frac{\sigma}{r} \right)^{12}
- \left( \frac{\sigma}{r} \right)^6 \right]   ,
\end{equation}

only two material parameters can be fitted. As the length parameter
$\sigma$ only sets the length scale, it is used to fit the lattice
constant $a$. Conventionally, the energy parameter $\eps$ is fitted to
the cohesive energy $\Ecoh$;\cite{Bru82} for the present work, however,
the elastic properties are more important, and we hence fit $\eps$ to
the bulk modulus $B$. Our fit parameters thus read: $\eps = 0.1515$ eV
and $\sigma=2.338$ \AA. Table~I proves that the cohesive energy is
severely underestimated in this potential, while the two elastic moduli
$c_{11}$ and $c_{12}$ are reproduced fairly well, within 12 \%. In the
conventional fitting scheme, which reproduces $\Ecoh$ by setting $\eps =
0.45$ eV, the bulk modulus -- and hence the elastic moduli in general --
are overestimated by a factor of 3.

We note that the LJ potential obeys a simple scaling property, which
allows us to extend the results obtained in the present study to other
systems by scaling lengths to $\sigma$ and energies to $\epsilon$. In
this sense the results obtained for the LJ potential are `universal'.

Both pair potentials are smoothly cut off at $\rcut=6.4$ \AA, i.e.,
after the 6th neighbour shell. We chose the following
procedure\cite{Vot93} -- here described for the LJ potential --

\begin{equation}  \label{potcut}
V(r)= \VLJ(r) - \VLJ(\rcut) + {\VLJ}'(\rcut) \frac{\rcut}{m}
\left[ 1 - \left( \frac{r}{\rcut} \right)^m \right]  ,
\end{equation}

where the prime $\VLJ'$ denotes differentiation with respect to $r$, and
a value of $m=20$ has been adopted. For the Morse potential, we proceed
analogously.

\subsection{Generalized stacking fault energy}

Dislocations in fcc metals consist of two partial dislocations between
which a stacking fault extends. The ability of potentials to describe
the formation and the energetics of a stacking fault is therefore
crucial when modelling plasticity. These features are conventionally
described with the help of the so-called `generalized stacking fault
energy' of the crystallographic (111) plane. It is defined as
follows.\cite{Vit68} Consider an ideal lattice with total energy $E_0$.
We cut it along a (111) plane into two halves. The upper half is shifted
parallel to the bounding (111) plane with respect to the lower half by a
vector

\begin{equation}
   \vec{f} = \alpha\vec{a}+\beta\vec{b} \qquad 0 \leq \alpha, \beta < 1
.
\end{equation}

Here the vectors $\vec{a}$ and $\vec{b}$ span the
(111)  surface:
$\vec{a}=\frac{1}{2}[11\bar{2}]$,
$\vec{b}=\frac{1}{2}[1\bar{1}0]$.

The generalized stacking fault energy is then defined in terms of the
energy $E(\alpha,\beta)$ of the distorted crystal as

\begin{equation} \label{GSFE}
   \gamma(\alpha,\beta) = E(\alpha,\beta)-E_0  .
\end{equation}

For calculating the energy of the distorted crystal, a
conjugate-gradient method is used; quenching will lead to the same
numerical results. All particles are constrained to move only in the
normal direction. It is crucial to choose the crystal sufficiently large
in the direction normal to the stacking-fault plane in order to obtain
stable solutions; unstable solutions lead to a discontinuous energy
surface. We chose a size of 10 unit cells in the lateral directions, and
25 unit cells for each half-crystal in normal direction.

In Fig.~1 we display the generalized stacking fault energy in \two\
direction. The values of the stable stacking fault energy $\gams$ and of
the unstable stacking fault energy $\gamu$ -- that is the energy barrier
between the undistorted lattice position and the stacking fault position
-- are also assembled in table~II. We note that the values for $\gamu$
predicted by the three potentials are not too far from each other; this
demonstrates that the value of the bulk modulus -- which has been chosen
identical in the three potentials -- has a major influence on the value
of $\gamu$. Quantitatively, the Morse (LJ) potential predicts a value of
$\gamu$ which is by 17 \% (31 \%) too high in comparison with the
prediction of the EAM potential; we note that no experimental value for
this quantity is available. The values of the stable stacking fault
energy $\gams$ vary quite dramatically from each other. We note that the
EAM potential predicts a value which is quite close to the experimental
value\cite{CR77} of 45 mJ/m$^2$. The values of $\gamu=5.9$ mJ/m$^2$ for
the Morse potential and of 10.8 mJ/m$^2$ for the LJ potential are
considerably too small. The fact that the stable stacking fault energy
is so small is not untypical of pair potentials. For the LJ potential,
for instance, it is well known that for infinite cutoff radius,
$\rcut=\infty$, the hcp phase has almost identical (in fact, even
smaller) energy as the fcc phase;\cite{BD55,WP65} even though
differences between the two phases become larger for finite cutoff
radius,\cite{JBA02} this fact makes a small value of $\gams$ plausible.

\subsection{Simulation}

We employ the method of classical molecular-dynamics to model the
indentation process. Our substrate consists of an fcc crystal. In the
case of a (100) surface, it has a side length of 70 lattice constants in
all directions and contains 1,372,000 atoms. In the case of the (111)
surface, similar crystal dimensions have been chosen, and the crystal
contains 1,381,800 atoms. We checked in a series of simulations that our
crystallite size is large enough to obtain reliable results for the
indentation process by investigating the force and pressure depth
curves, atomistic snapshots and the defect dynamics.

Lateral periodic boundary conditions have been applied. At the bottom,
atoms in a layer of the width $\rcut$ have been constrained to $F_{\rm
normal} = 0$; we checked that increasing the width of this layer to
$(2\dots3)\rcut$ -- as it is appropriate for the EAM potential -- has no
influence on the results of our indentation simulations. The simulations
have been done in the microcanonical ensemble at $T=0$ K using a
modified version of the LAMMPS code.\cite{LAMMPS}

We found that a careful relaxation of the crystal before starting the
indentation process is crucial in order to obtain reliable and
reproducible results. The substrate has been relaxed to $p_{ij} <
10^{-5}$ GPa and temperatures $\ll 1$ K using a very low frictional
force and pressure relaxation in a Nose-Hoover algorithm. Upon
incomplete relaxation, we encountered artefacts like oscillations in the
response functions during the indentation process and an overestimation
of the material strength; these features are caused by the remaining
internal stress fields in the crystal.

The indenter is modelled as a repulsive soft sphere. We chose a
non-atomistic representation of the indenter, since we are not
interested in the present study in any atomistic displacement processes
occurring in the indenter, but only in the substrate. The interaction
potential between the indenter and the substrate atoms is limited to
distances $r<R$, where $R$ is the `indenter radius'. At $r<R$, the
potential smoothly increases like\cite{KPH98}

\begin{equation} \label{5}
V(r) = \left\{ \begin{array}{ll}
k(R-r)^3, & r<R, \\
0, & r\ge R .
\end{array} \right.
\end{equation}

We call $k$ the \emph{indenter stiffness}. For the results presented
here, it has been set to $k=3$ \eVA. We checked that our results are
only weakly influenced by the exact value of the contact stiffness, as
long as it is in the range of $1-10$ \eVA. Only when decreasing $k$ to
below 0.1 \eVA, the results change sensitively; this can be understood
since the decreased indenter stiffness translates into a smaller
indenter `elastic modulus'.

Our indenter has a radius of $R=8$ nm. This value was chosen as a
compromise; for larger indentation radii, the influence of the finite
size of the simulation volume shows up, while with decreasing $R$, the
atomistic nature of the indentation process leads to increased
fluctuations, in particular in the contact area. We made sure in a
series of simulations, that in the range of $R=3-15$ nm in the elastic
regime, no systematic deviations from the Hertzian theory appear.

We tested two methods of indentation, a \emph{displacement-controlled}
and a \emph{velocity-controlled} approach. When controlling the
displacement, we advance the indenter every $\Dt=2$ ps by a fixed amount
of $\delta= 0.256$ \AA\ ($\ll$ lattice constant) instantaneously,
corresponding to an average indentation speed of $v=12.8$ m/s. The
substrate then relaxes for the ensuing time of $\Dt$ to the new indenter
position. In the velocity-controlled method, a fixed indentation speed
of $v=12.8$ m/s is imposed on the indenter. We found no systematic
differences in the material response nor in the induced plasticity
between the simulation results obtained by the two methods. We prefer to
use the former method. We note that in experiment, either displacement
or force can be controlled.\cite{Fis04} In the following all ensemble
properties are obtained by averaging over the $2$ ps relaxation cycle.
All our simulations were performed without damping; the resulting energy
input into the crystal amounted to less than 6 meV/atom.

\section{Results} \label{Results}

\subsection{Elastic regime: load-displacement curves}

In Fig.~2 we display the basic information obtained from nanoindentation
simulation, the force-displacement curves. For the two crystal surfaces
studied, the (100) and the (111) surface, the three potentials give
results which are qualitatively in agreement with each other. However,
in detail deviations are visible which will be discussed in the
following.

Let us first look at the elastic regime which is well described by the
Hertzian law,\cite{Her82,Fis04}

\begin{equation} \label{Hertz}
 F = \frac43 \Er d^{3/2} \sqrt{R}.
\end{equation}

Here, $\Er$ is the so-called indentation modulus,
which for an isotropic solid can be expressed as

\begin{equation} \label{Eriso}
\Er = \frac{E}{1-\nu^2}
\end{equation}

in terms of the Young modulus $E$ and the Poisson ratio $\nu$ of the
material. However, the material discussed here is strongly anisotropic.
The anisotropy of cubic materials can be expressed in terms of the
elastic moduli via

\begin{equation} \label{ani}
\ani = \frac{2c_{44}}{c_{11} - c_{12}} ,
\end{equation}

Its value is presented in Table~III. Due to the fact that the pair
potentials do not model all three elastic constants, they predict wrong
values for the anisotropy; however, in all potential models discussed
here, Cu is strongly anisotropic.

Vlassak and Nix\cite{VN93,VN94} have shown that in an anisotropic solid,
the Hertzian force-displacement law \queq{Hertz} remains valid if an
orientation-dependent indentation modulus $\Er(\Om)$ is appropriately
chosen; they give numerical tables to evaluate $\Er(\Om)$ as a function
of the elastic constants. We collect the results evaluated for our
potentials in Table~III. It is seen that the (111) surface is stiffer
than the (100) surface. As expected, the EAM results are in close
agreement with the moduli calculated using the experimental data. For
the (100) direction, the Morse potential yields indentation moduli which
fairly well reproduce those of the EAM potential, while the LJ potential
is considerably -- by 41\% -- too stiff. For the (111) surface also the
Morse prediction overestimates (by 40 \%) the indentation modulus; this
is due to the fact that the anisotropy $A$, which is only poorly
represented by the pair potentials, sensitively enters the modulus in
this direction.

In table III, we also display fit values $\Erfit$ for the indentation
moduli, which have been obtained by fitting Fig.~2 to the Hertzian law,
\qeq{Hertz}. We perform this fit only for the initial part of the
indentation curve ($d<2$ \AA) in order to stay in the regime of linear
elasticity. When we compare with the indentation moduli, we see an
overall fair agreement with at most 12 \% deviation. We attribute these
minor deviations to (i) the atomistic nature of the indentation process,
which for the indenter radius of 8 nm is not fully captured by continuum
elasticity; (ii) the possible onset of nonlinear elastic processes -- as
these tend to make the material respond more stiffly to the applied
force,\cite{ZLV*04} this would agree with our finding that most fitted
moduli are larger than the theoretical prediction;
(iii) the numerical problem of fitting the
molecular-dynamics data to the Hertzian law -- note that besides the
elastic deformation also a finite offset in the displacement has to be
fitted. We conclude that the Vlassak-Nix indentation moduli give an at
least semi-quantitatively correct representation of the elastic behavior
of the three potentials, and allows us to understand the trends in the
elastic part of the force-displacement curves in Fig.~2.

\subsection{Onset of plasticity}

In order to discuss the onset of plasticity and hardness, the evolution
of the contact pressure is displayed in Fig.~3. These data are obtained
from the forces $F$ of Fig.~2 by dividing through the pertinent
projected contact areas. For the (111) surface, the critical indentation
depth $\dy$ can be quite clearly identified at around $8-10$ \AA; after
this indentation, plasticity sets in (see also Fig.~10 below). The
contact pressure at $\dy$ is called the critical stress $\sigcr$. When
$\dy$ is exceeded, a quite considerable load drop is experienced for the
(111) surface, which corresponds to a softening of the material due to
the production of mobile defects. For the (100) surface, the onset of
plasticity is not as sharp and already starts at around 5 \AA, cf.\ also
Fig.~10 discussed below.

These principal features of the two surfaces are qualitatively well
reproduced by all three potentials. The origin of the abrupt onset of
plasticity for the (111) surface is twofold: (i) the primary glide
systems \slip\ are located at quite oblique angles to the direction of
the indentation force acting normally to the surface; the corresponding
Schmid factor is only $\sone = \sqrt{2/27} = 0.27$. For the (100)
surface these glide systems are more easily activated, since $\szero =
1.5 \sone= 1/\sqrt{6}=0.41$. (ii) When finally the critical indentation
depth has been reached, a considerable elastic energy has built up due
to the high stiffness of this surface. Then, upon dislocation
nucleation, a stronger dislocation avalanche and consequently a larger
plastic displacement jump are achieved.

The values of the critical stresses obtained in the simulation are
assembled in table~IV. At these stresses, the resolved shear stress on
the slip plane exceeds the theoretical shear strength -- that is the
maximum shear stress that a defect-free solid can sustain without
yielding -- of the crystal. The theoretical strength $\tauth$ of an fcc
single crystal has been calculated by Frenkel\cite{Fre26,KM86} as

\begin{equation}  \label{Frenkel}
\tauth = \frac{1}{\sqrt{2}} \frac{1}{2\pi} \Gr ,
\end{equation}

where $\Gr$ is the (so-called `relaxed')\cite{KM86,RKCM99} shear modulus
for the preferred glide system \slip\ that is calculated from the
elastic constants as\cite{Fre26,KM86}

\begin{equation}  \label{shear}
\Gr = \frac{3 c_{44} (c_{11} - c_{12} ) }
      { 4 c_{44} +c_{11} - c_{12} } .
\end{equation}

In \qeq{Frenkel} we also take the fact into account that the interplanar
distance of the glide planes is larger than the partial Burgers vector
by a factor of $1/\sqrt{2}$.\cite{KM86}

Recent ab initio calculations of the theoretical strength of Cu (and
other fcc metals) indicate that Frenkel's estimate \queq{Frenkel} is
quite accurate; they only substitute the prefactor $1/\sqrt{2}2\pi =
0.11$ by 0.085;\cite{RKCM99,OLY02,OLH*04} for the present purposes it
will be sufficient to stay with Frenkel's original estimate
\queq{Frenkel}. The values of $\Gr$ and $\tauth$ predicted by the
potentials are included in table~IV.

For an anisotropic crystal, it is a nontrivial task to establish a
quantitative relationship between the theoretical strength $\tauth$ of
the solid -- that is the maximum shear stress that a defect-free solid
can sustain without yielding -- and the critical stress $\sigcr$, as it
is applied in a nanoindentation experiment on the surface of the
crystal. For isotropic solids, this relationship has been analyzed and
it has been found that for a material with a Poisson ratio of 0.3, the
maximum shear stress occurs along the axis of symmetry at a depth of
approximately $a/2$, where $a$ is the radius of the contact
area.\cite{Tab51,Joh85book,Fis07} At this point, the shear stress is
given by

\begin{equation}  \label{stress}
\tau = 0.47 \sigma   ,
\end{equation}

where $\sigma$ is the contact pressure, i.e., the mean stress on the
surface. This maximum shear acts on planes inclined to the surface at
$45^\circ$. Since this resembles the situation of a solid under uniaxial
stress, the Schmid factor might be expected to provide a valid though
rough estimate of the maximum resolved shear stress on the slip planes
underneath the indenter. This reasoning would lead to the prediction
that the critical stresses should behave in inverse proportion to the
relevant Schmid factors

\begin{equation}  \label{Schmid}
\frac{\sigcr(111)}{\sigcr(100)} = \frac{\szero}{\sone} = 1.5.
\end{equation}

As Table IV shows, indeed $\sigcr(111) > \sigcr(100)$, but the ratio is
only around $1.2 - 1.3$, rather than 1.5. Since the Schmid factors
simply express the geometric orientation of the slip systems with
respect to the surface plane, the disagreement of the molecular-dynamics
results with the prediction \queq{Schmid} indicates that the stress
state inside an anisotropic material looks rather different from the
above simple reasoning. If we nevertheless attempt to use the isotropic
result \queq{stress} to connect the external average stress on the
surface to the internally active shear stress, and naively include the
Schmid factor to take the surface orientation into account, we arrive at

\begin{equation}  \label{taunew}
\sigcr = \frac{1}{0.47s} \tauth =
 \left\{ \begin{array}{ll}
5.2 \tauth, & (100), \\
7.9 \tauth, & (111)  .
\end{array} \right.
\end{equation}

A comparison with table IV shows that this result allows to explain
satisfactorily the order of magnitude of the critical stresses observed.
When we consider furthermore the value of the critical stresses for the
different potentials, we see that the critical stress of the LJ
potential is highest (this fact agrees with the overestimation of the
elastic properties), while those of the Morse and the EAM potentials are
quite similar, despite the difference in the unstable stacking fault
energies in these potentials.

The critical indentation depths for the different potentials show the
inverse behavior. Since the contact pressure rises strongest for the LJ
potentials the critical indentation depth is the smallest, while for
Morse and EAM potential it has almost the same value. Overall, the
differences in the critical indentation depths are not as pronounced as
those for the critical stresses.

For the (100) surface, the discussion of the critical indentation depth
is not useful due to the very smooth onset of dislocation nucleation. We
note, however, that at a depth of around $8-10$ \AA, also for this
surface the maximum stress is observed for the EAM and LJ potentials,
which is of similar size as the critical stress in the (111) surface.
Upon further indentation, however, no load drop is experienced, but only
a drop in contact pressure. This feature is due to the considerable
increase in dislocations observed at this point, see also Fig.~10 below.

\subsection{Hardness}

The contact pressure which has been established after the critical
indentation depth has been exceeded and a possible load drop has
occurred, stays rather constant with increasing indentation; this
defines the hardness of the material. Fluctuations in this regime are
due to the atomistic resolution of the processes and immediately reflect
the generation and interaction of dislocations with each other. Note
that the hardness for the EAM potential remains rather constant, while
that of the pair potentials tends to increase somewhat with increasing
indentation depth. This feature is particularly pronounced for the LJ
potential and reflects the work hardening of the material due to the
high density of partial dislocations with extended stacking faults, cf.\
\qsect{snapshots} below.

\subsection{Atomistic Snapshots}  \label{snapshots}

We present in Figs.~4 -- 9 atomistic views of the defects created in the
material during the indentation process. These views have been obtained
at different penetration depths: (i) 'Embryonic plasticity', obtained
immediately at the critical indentation depth, Figs.~4 and 6; (ii)
`emerging plasticity', where the dislocations formed are clearly
discernible, but are still, more or less, localized under the indenter;
and (iii) 'fully-developed plasticity', where the dislocations have
started to glide away from their point of production and fill the
simulation volume. The plots have been generated using the recently
developed algorithm by Ackland and Jones.\cite{AJ06} This algorithm
classifies all atoms according to their atomic neighborhood. In our
case, we have fcc atoms (these constitute the vast majority of atoms and
are not displayed for clarity), surface atoms (grey), and stacking fault
atoms with a local hcp structure (red). All other atoms are categorized
as atoms of lower symmetry and plotted in green. The boundaries of
dislocations, positions where dislocations interact, and also embryonic
defects are shown in this color.

Fig.~4 captures embryonic plasticity for the (100) surface just at the
stage of homogeneous dislocation nucleation. For the Morse and EAM
potential, the nucleating defect structures have not even formed
stacking faults that could have been recognized by the detection
algorithm. In the LJ potential, nucleation occurred somewhat earlier but
at higher stresses than in the EAM case, cf.\ Fig.~10 discussed below.
In all cases the nucleation process is seen to start homogeneously under
the indenter, viz.\ in the region of highest shear stress on the (111)
slip planes with highest resolved shear stress. For the (111) surface,
Fig.~5, an instance of time has been selected where the nucleated
dislocations are clearly discernible by their stacking fault planes. A
comparison of Fig.~4 and 5 demonstrates that indeed the LJ system
develops the largest stacking fault planes. This is connected to the
extremely low stable stacking fault energy $\gams$ of the LJ potential,
which allows partial dislocations to propagate more easily.

The indentation depths of Figs.~6 and 7 have been chosen such that the
dislocations are still
localized in the region of highest shear stress under the indenter.
Here, again the unrealistically large stacking fault planes obtained for
the LJ potential (Fig.~6a) are seen; in Fig.~7b also the Morse potential
has formed large stacking faults. Note, however, that for the (111)
surface the defect structures for the three potentials look more similar
to each other. Characteristically, for the realistic EAM potential, the
emission of a prismatic loop is already seen at this stage, Fig.~6c.
This already points at an easier possibility for cross slip occurring
under this potential and is connected to the fact that work hardening
occurs in the defective region. This feature will show up again in the
case of fully developed plasticity.

For fully developed plasticity, Figs.~8 and 9, the emission of prismatic
loops is observed for all potentials. For the LJ potential depicted in
Fig.~9a a strikingly large number of prismatic loops is emitted. Again,
dislocations in the LJ system are characterized by huge stacking faults.
In the case of the pair potentials, in addition an exceptionally high
number of V-shaped dislocations at the surface an be seen: a small one
in Fig.~8a, and huge ones in Fig.~9b. These dislocations are prismatic
loops moving parallel to the surface, and contain a higher core energy
density than those in bulk material.

In summary, we can rationalize our atomistic results on the development
of plasticity for the potentials investigated here using two concepts:
(i) The extremely low stable stacking fault energy of pair potentials,
in particular for the LJ potential, allows for the formation of large
stacking fault planes in these systems. (ii) In the realistic EAM
potential, on the other side, dislocations are more compact, have
smaller dissociation widths, interact less with each other and have a
higher chance of cross-slipping, leading to less work hardening in this
system.

In Fig.~10 we quantify the amount of dislocations formed. To this end,
we plot the fraction of atoms classified as stacking fault atoms
(colored red in Figs.~4 -- 6). Consistently with our atomistic
snapshots, the number of stacking faults formed is considerably smaller
for the realistic EAM potential than for the pair potentials.
Furthermore, the LJ potential exhibits the highest number of stacking
faults. This plot thus characterizes the influence of the stacking fault
energy $\gams$ on the amount of plasticity formed.

\section{Conclusions}

We investigate the influence of the form of the interatomic potential on
the emergence of plasticity in nanoindentation. Specifically, we model
indentation into a Cu single crystal. Two pair potentials, LJ and Morse,
are compared with an EAM many-body potential. We took care that in all
potentials, the lattice constant -- as a measure of the atomistic depth
scale in nanoindentation -- and the bulk modulus -- as a measure of the
elastic stiffness of the material -- take identical values. We find:

\begin{enumerate}

\item In zeroth order, the force-displacement curves obtained for the
three potentials coincide surprisingly well. We find that qualitatively
many aspects of nanoindentation are fairly well reproduced by simple
pair potentials. This applies to the elastic Hertzian deformation, the
onset of plasticity, and the gross value of the hardness. This general
qualitative agreement makes pair potentials useful for parameter and
sensitivity studies.

\item Among the correctly represented features is also the influence of
the crystal orientation; even though the numerical value of the crystal
anisotropy is not modelled exactly, pair potentials correctly predict
the fact that the load drop is larger for the (111) than for the (100)
surface, as well as the location of the critical indentation depth.

\item However, in detail, important quantitative deviations appear,
which can be traced back to the potentials used.

\begin{enumerate}

\item Pair potentials -- and here in particular the LJ potential -- fail
in giving a quantitative representation of the elastic part of the
indentation curve; this is due to the fact that pair potentials are in
principle unable to model all three elastic moduli of a cubic crystal;
in the case of a LJ potential, only one elastic constant can be
modelled.

\item As a consequence, the elastic anisotropy is wrongly modelled in
pair potentials; thus, the elastic response of the various crystal
orientations is not correctly reproduced. Pair potentials which allow
for the independent representation of two elastic constants (like the
Morse potentials) fare better than those in which only one elastic
constant is correctly modelled (like the LJ potentials).

\item Since the theoretical strength of a material can be expressed
approximately as a linear function of the elastic constants, the pair
potentials predict quantitatively slightly wrong values of the
theoretical strength and hence the hardness of the material.

\item Similarly the critical stress, i.e., the contact pressure at the
critical indentation depth, is wrongly predicted by the pair potentials;
it is overestimated in our case.

\end{enumerate}

\item As a major result we could identify the influence of the stable
and unstable stacking fault energies on the plasticity and dislocation
activity in the plastic regime. In the cases investigated, the unstable
stacking fault energy roughly coincided (within 30 \%) for the three
potentials; this correlates well with the fact that the critical stress
for plastic  yielding was roughly similar. However, the stable
stacking fault energy differed by an order of magnitude; in particular,
the pair potentials had a considerably lower stable stacking fault
energy than that predicted by the EAM potential (which is close to
experimental data). As a result, dislocations simulated by these pair
potentials tend to have large stacking faults, exhibit a faster
expansion of partials and consequently a stronger dislocation
interaction, resulting in a stronger work hardening. In contrast, the
EAM potential with its more realistic stacking fault energy leads to
earlier generation of prismatic loops, easier cross slip of dislocations
and less work hardening. Thus, in particular the modelling of fully
developed plasticity will show unrealistic features when modelled using
pair potentials.

\end{enumerate}

\begin{acknowledgments}

The authors acknowledge financial support by the Deutsche
Forschungsgemeinschaft via the Graduiertenkolleg 814, and a generous
grant of computation time from the ITWM, Kaiserslautern.

\end{acknowledgments}

\newpage

\begin{table}

\begin{tabular}{l||l|l|l|l|l|l}
Potential      & $\Ecoh$ (eV) & $a$ (\AA) &
$B$(GPa)  & $c_{11}$ (GPa) & $c_{12}$ (GPa) &
$c_{44}$ (GPa) \\
\hline
LJ         & 1.19    & 3.615* & 138.2*  & 193.5  & 110.5   & $=c_{12}$ \\
Morse      & 3.54*   & 3.615* & 138.2*  & 172.3  & 121.2   & $=c_{12}$ \\
EAM        & 3.54*   & 3.615* & 138.4*  & 169.9* & 122.6*  & 76.2* \\
\hline
Experiment & 3.54    & 3.615  & 138.3   & 170.0  & 122.5   & 75.8
\end{tabular}

\caption{ Lattice properties of Cu as represented by the potentials
employed in comparison with experimental data.
The values marked with an asterisk have been used
for fitting the respective potential. Experimental data
taken from \qref{MMP*01} and the references quoted therein.}

\label{t1}
\end{table}

\begin{table}

   \begin{tabular}{l|l|l}
      Potential &  $\gamu$ (mJ/m$^2$) &   $\gams$ (mJ/m$^2$) \\
      \hline
      LJ    & 229.2   & 10.8   \\
      Morse & 204.0   & 5.9    \\
      EAM   & 174.4   & 43.3
   \end{tabular}

   \caption{Stable and unstable stacking fault energies.
}

\label{t2}
\end{table}

\begin{table}

    \begin{tabular}{l||l|l|l||l|l}
      Potential & $\ani$ & $\Er(100)$ (GPa) & $\Erfit(100)$ (GPa)
      & $\Er(111)$  (GPa)  & $\Erfit(111)$  (GPa)
       \\
      \hline
      LJ            & 2.66   & 185.9  & 185  & 216.0   & 235 \\
      Morse         & 4.74   & 157.6  & 145  & 203.0   & 195 \\
      EAM           & 3.22   & 135.0  & 134  & 151.9   & 171 \\ \hline
      experiment    & 3.19   & 131.5  & $-$  & 153.8   & $-$
      \end{tabular}

   \caption{Anisotropy $\ani$, orientation-dependent indentation modulus
$\Er$ predicted by the theory of Vlassak and Nix\cite{VN93,VN94}
and the moduli $\Erfit$, as obtained from a fit of the
force-displacement data of  Fig.~2 to the Hertzian law \queq{Hertz}.
}

\label{t3}
\end{table}

\begin{table}

\begin{tabular}{l||l|l||l|l||l|l|l}
Potential  &  $\Gr$ (GPa) &  $\tauth$ (GPa)
 &  $\sigcr(100)$ (GPa)   & $\sigcr(111)$ (GPa)
 &  $\sigcr(100)/\tauth$  & $\sigcr(111)/\tauth$
 & $\sigcr(111)/\sigcr(100)$ \\
\hline
LJ         & 52.4  & 5.90   & 23.9   & 27.9 & 4.1 &  4.8  & 1.17 \\
Morse      & 34.7  & 3.90   & 18.8   & 23.5 & 4.8 &  6.0  & 1.25 \\
EAM        & 30.7  & 3.46   & 16.5   & 22.0 & 4.8 &  6.4  & 1.33 \\
\hline
Experiment & 30.8  & 3.47   & $- $   & $-$  & $-$ &  $-$  & $-$

\end{tabular}

\caption{
Shear moduli $\Gr$, \qeq{shear}, theoretical strengths $\tauth$,
\qeq{Frenkel}, and critical shear stresses $\sigcr$ taken from Fig.~3.}

\label{t4}
\end{table}

\newpage \clearpage

\begin{figure}

   \includegraphics[width=0.45\textwidth]{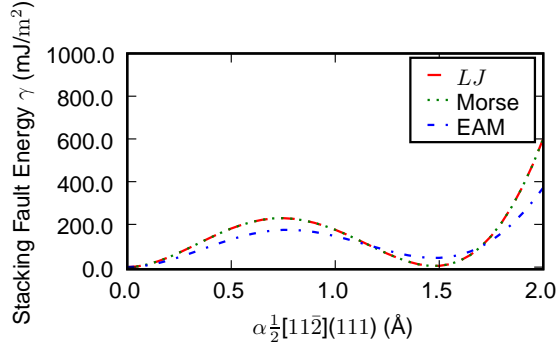}

   \caption{Generalized stacking fault energy $\gamma$ along
the \two\ direction.}

\label{f1}
\end{figure}

\begin{figure}

\subfigure[]{\includegraphics[width=0.45\textwidth]{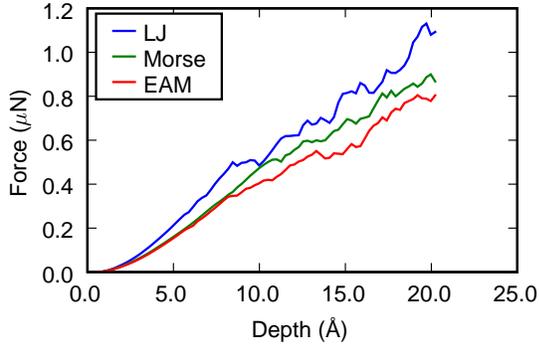}} \hfill
\subfigure[]{\includegraphics[width=0.45\textwidth]{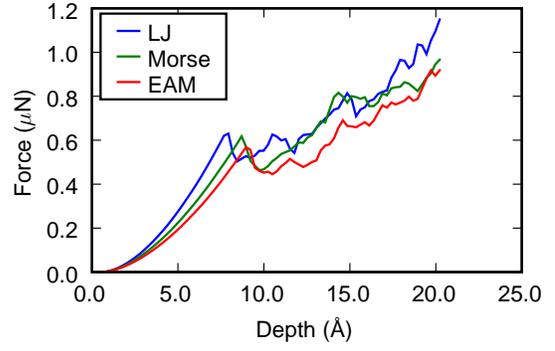}}

\caption{Force $F$ vs displacement $d$ for Cu. Crystal orientation: a)
(100) b) (111). }

   \label{f2}
\end{figure}

\begin{figure}

\subfigure[]{\includegraphics[width=0.45\textwidth]{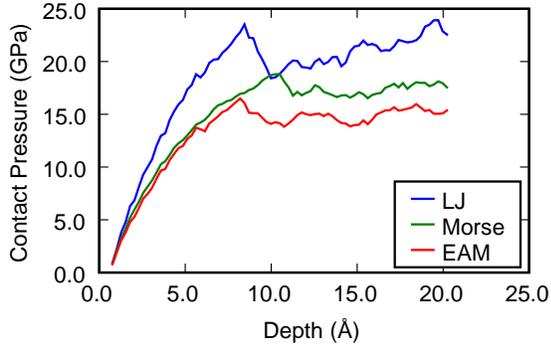}}  \hfill
\subfigure[]{\includegraphics[width=0.45\textwidth]{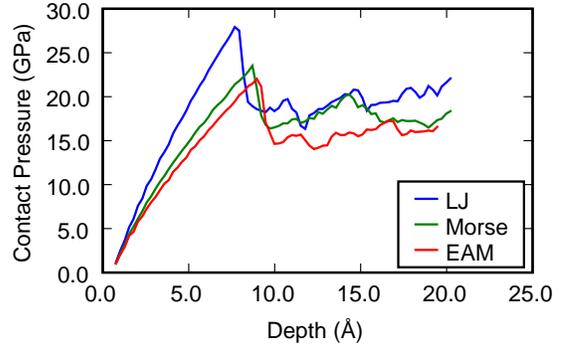}}

\caption{Contact pressure vs displacement $d$ for Cu. Crystal orientation:
a) (100)
b) (111). }

\label{f3}
\end{figure}

\begin{figure}

\subfigure[]{\includegraphics[width=0.45\textwidth]{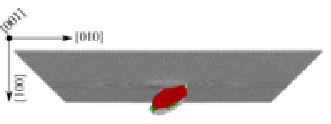}} \hfill
\subfigure[]{\includegraphics[width=0.45\textwidth]{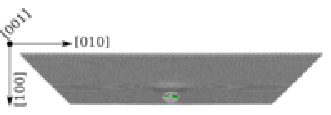}}   \\
\subfigure[]{\includegraphics[width=0.45\textwidth]{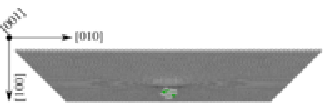}}

   \caption{ Snapshots of embryonic plasticity ($d=5.6$ \AA) for
(100) Cu. a) LJ b) Morse c) EAM potential. }

   \label{f4}
\end{figure}

\begin{figure}

\subfigure[]{\includegraphics[width=0.45\textwidth]{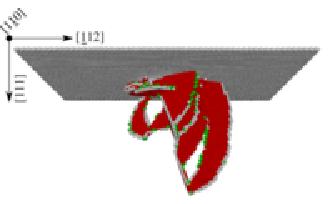}} \hfill
\subfigure[]{\includegraphics[width=0.45\textwidth]{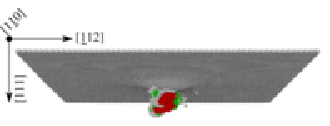}}   \\
\subfigure[]{\includegraphics[width=0.45\textwidth]{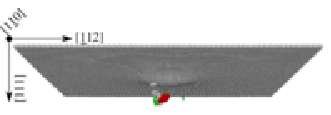}}

   \caption{ Snapshots of embryonic plasticity ($d=8.7$ \AA) for
(111) Cu. a) LJ b) Morse c) EAM potential. }

   \label{f5}
\end{figure}

\begin{figure}

\subfigure[]{\includegraphics[width=0.45\textwidth]{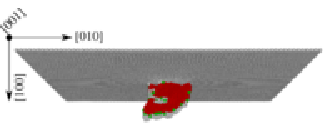}}  \hfill
\subfigure[]{\includegraphics[width=0.45\textwidth]{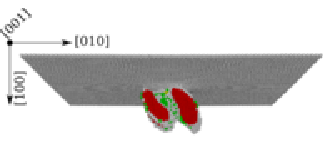}}    \\
\subfigure[]{\includegraphics[width=0.45\textwidth]{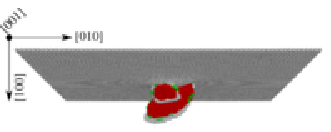}}

   \caption{ Snapshots of emerging plasticity ($d=8.2$ \AA) for
(100) Cu. a) LJ b) Morse c) EAM potential. }

   \label{f6}
\end{figure}

\begin{figure}

\subfigure[]{\includegraphics[width=0.45\textwidth]{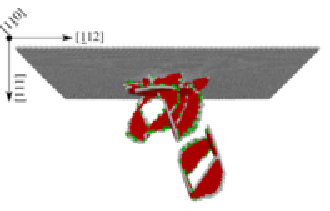}}   \hfill
\subfigure[]{\includegraphics[width=0.45\textwidth]{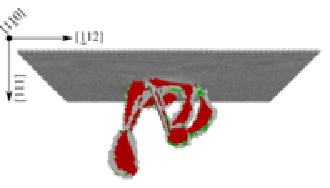}}   \\
\subfigure[]{\includegraphics[width=0.45\textwidth]{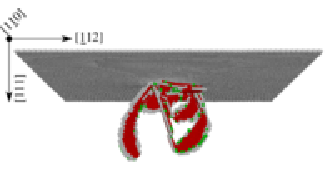}}

   \caption{ Snapshots of emerging plasticity ($d=9$ \AA) for (111) Cu.
a) LJ b) Morse c) EAM potential. }

   \label{f7}
\end{figure}

\begin{figure}

\subfigure[]{\includegraphics[width=0.45\textwidth]{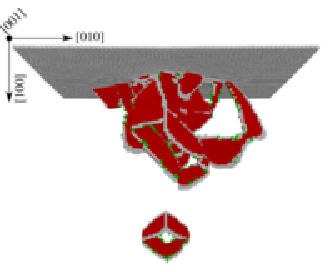}}  \hfill
\subfigure[]{\includegraphics[width=0.45\textwidth]{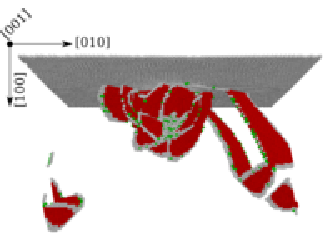}}    \\
\subfigure[]{\includegraphics[width=0.45\textwidth]{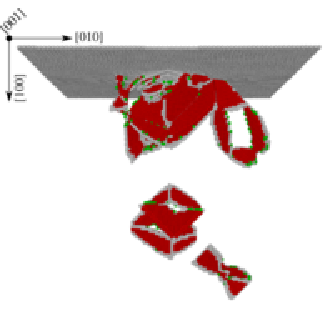}}

   \caption{ Snapshots of fully developed plasticity ($d=14.1$ \AA) for
(100) Cu. a) LJ b) Morse c) EAM potential. }

   \label{f8}
\end{figure}

\begin{figure}

\subfigure[]{\includegraphics[width=0.45\textwidth]{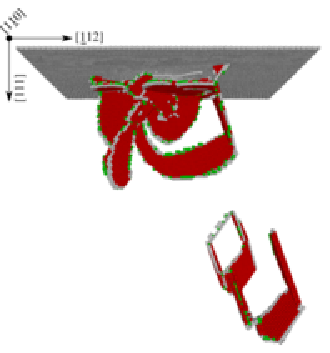}} \hfill
\subfigure[]{\includegraphics[width=0.45\textwidth]{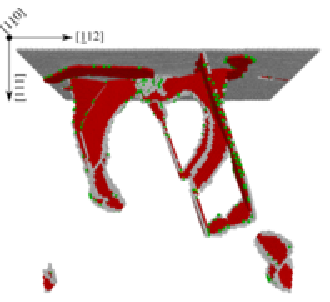}}   \\
\subfigure[]{\includegraphics[width=0.45\textwidth]{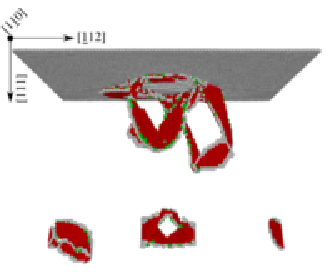}}

   \caption{ Snapshots of fully developed plasticity ($d=14.1$ \AA) for
(111) Cu. a) LJ b) Morse c) EAM potential. }

   \label{f9}
\end{figure}

\begin{figure}

\subfigure[]{\includegraphics[width=0.45\textwidth]{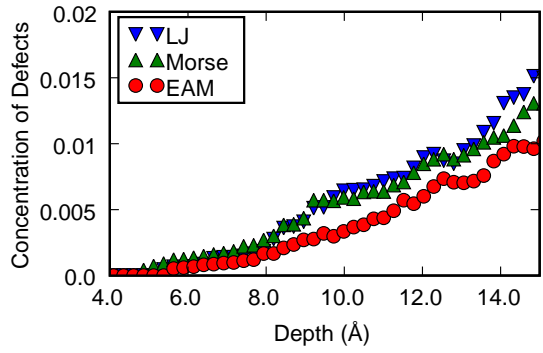}}  \hfill
\subfigure[]{\includegraphics[width=0.45\textwidth]{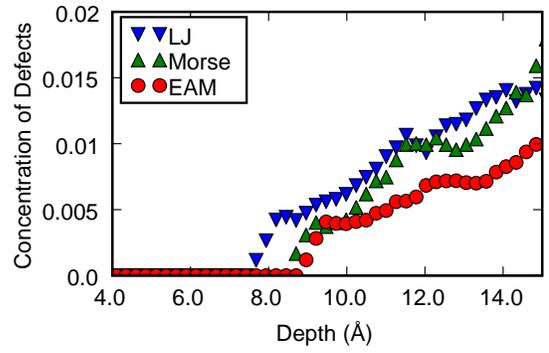}}

\caption{Concentration of Stacking Faults, i.e., fraction of atoms
occupying stacking-fault sites. Crystal orientation: a) (100) b) (111).}

\label{f10}
\end{figure}

\clearpage

\end{document}